\documentclass[12pt]{article}
 \usepackage{amsfonts}
 \usepackage{amssymb,amsmath}
 \usepackage{graphicx} \usepackage{epstopdf}
 \newcommand{\crlb}[1]{\label{#1}\\[2pt]}
 
 \newcommand{\crld}[1]{\label{#1}}
 \newcommand{\eela}[1]{\quad\hbox{\scriptsize{#1}}\label{#1}\end{eqnarray}}
 \newcommand{\eelb}[1]{\label{#1}\end{eqnarray}}
 
 \newcommand{\newsecb}[2]{\section{#1}\label{#2}\setcounter{equation}{0}}

 \newcommand{\nolabels} {\def\eel{\eelb}\def\eeql{\eeqlb}  \def\crl{\crlb} 
 \def\newsecl{\newsecb}\def\bibiteml{\bibitem} \def\citel{\cite}\def\labell{\crld}}
\newcommand{\eeqla}[1]{\quad\hbox{\scriptsize{#1}}\label{#1}\end{aligned}\end{equation}}
\newcommand{\eeqlb}[1]{\label{#1}\end{aligned}\end{equation}}

\newcommand\publishversion  {\nolabels\setlength{\textheight}{8.38in}\setlength
    {\oddsidemargin}{0in} \setlength{\textwidth}{6.2in}\setlength{\topmargin}{-0.2in}}
\def\beq{\begin{equation}\begin{aligned}}		\def\eeq{\end{aligned}\end{equation}}
\def\be{\begin{eqnarray}}  					\def\ee{\end{eqnarray}}		
   \def\bi#1{\begin{itemize}\item[#1]} 	     \def\itm#1{\item[#1]} 	   \def\ei{\end{itemize}} 
   \def\eqn#1{(\ref{#1})}
   	 \def\fn{\footnote}	 

		 \def\del{\delta}  % \def\k{\kappa}     \def\l{\lambda}  
 \def\alf{\alpha}
 \def\del{\delta}        \def\eps{\varepsilon} 
             
             \def\vv{\varphi}    
                  \def\sig{\sigma}

 \def\w{\omega}

  \def\ra{\rightarrow} 
  
 \def\dd{{\rm d}}  \def\bra{\langle}   \def\ket{\rangle}
				\def\deff{\overset{\mathrm{def}}{=}}

\def\fract#1#2{{\textstyle\frac{#1}{#2}}}	 	 	
\def\ffract#1#2{\raise .2 em\hbox{$\scriptstyle#1\,$}\kern-.34 em/\kern-.34 em\lower .15 em \hbox{$\scriptstyle\,#2$}}
\def\half{\fract12}					
 
\def\ex#1{e^{\textstyle#1}} 		\def\qqquad{\qquad\qquad}	
\def\bpmatrix{\begin{pmatrix}} 			\def\epmatrix{\end{pmatrix}}
\def\bmatrix{\begin{matrix}} 			\def\ematrix{\end{matrix}} 
\def\bcenter{\begin{center}}			\def\ecenter{\end{center}}

\def\lowerheightgth#1#2#3{\(\raise-#1\hbox{\includegraphics[height=#2]{#3}}\)}
\def\lowerwidthgth#1#2#3{\(\raise-#1\hbox{\includegraphics[width=#2]{#3}}\)}
\def\widthfig#1#2{\includegraphics[width=#1]{#2}}
\def\th{\({}^{\mathrm{th}}\,\)}		 
\def\intt{{\mathrm{int}}}    
\def\ontt{{\mathrm{ont}}}

\def\weglaten#1{}	
 
\publishversion % do not remove, see \newcommands above
 
\begin{document}
\begin{titlepage} 
\title{
Quantum Foundations as a Guide for Refining Particle Theories\fn{Submitted to the Proceedings of the 19\th Rencontres du Vietnam, 2023, ``Windows in the Universe".}
\author{Gerard 't~Hooft}}
\date{\normalsize
Faculty of Science,
Department of Physics\\
Institute for Theoretical Physics\\
Princetonplein 5,
3584 CC Utrecht \\
\underline{The Netherlands}\\
http://www.staff.science.uu.nl/\~{}hooft101 \\[-15pt]
}
 \maketitle 

\begin{quotation} \noindent {\large\bf Abstract } % \\[10pt]

All quantum field theories that describe interacting bosonic elementary particles, share the feature that the zeroth order perturbation expansion describes non-interacting harmonic oscillators. This is explained in the paper. We then indicate that introducing interactions still leads to classical theories that can be compared with the quantum theories, but only if we terminate the expansion somewhere.
`Quantum effects'  typically occur when some of the classical variables fluctuate too rapidly
to allow a conventional description, so that these are described exclusively in terms of their energy eigen modes; these do not commute with the standard classical variables.
Perturbation expansions are not fundamentally required in classical theories, and this is why they are more precisely defined than the quantum theories. Since the expansion parameters involve the fundamental constants of nature, such as the finestructure constant, we suggest that research in these classical models may lead to new clues concerning the origin of these constants.

 \end{quotation}\end{titlepage}
 
 \newsecl{Introduction}{intro.sec}  \setcounter{page}{2}

There are numerous formal arguments and philosophies that indicate the fundamental impossibility to mimic quantum mechanical behaviour in classical models. The present work suggests that the reason for this curious situation in physics is not that classical models are to be considered as outlawed, but that quantum and classical theories can only be compared in terms of their perturbation expansions. Overlooking the fact that such expansions diverge must lead to apparent failures of pure logic. Here we list a class of classical models that do allow a description in quantum mechanical terms, just because the perturbation expansion is trivial: the quantum harmonic oscillators.

The Standard Model of the sub-atomic particles is often presented as a theoretical description of all known particles that only has two basic shortcomings. First, the gravitational force cannot be added without generating uncertainties in the renormalisation procedure, and secondly, it seems that the gravitational effects observed in galaxies and groups of galaxies cannot be accounted for in terms of all particles that are observed;  something basic, possibly invisible and unknown forms of matter,  is missing. 

Naturally, it is attempted either to add as yet unknown particles to the set of elementary particles known, or to add new terms to the equations, in particular those of the gravitational force.

But if history of science has taught us one thing, it is that guessing does not often provide for the correct answers, and that the best procedure for improving our understanding consists of systematic studies of imperfections that can easily have been overlooked.

In this paper, we advocate to combine two logical weaknesses that, each in their own way have not provided for clues, but maybe  they will, when taken together:
\bi{One,} the formalism that produced the successes of he Standard Model, required us to take a logical step for granted, even if the mathematical basis is weakly founded: we have to apply perturbation expansions. If all scattering amplitudes are expressed as coefficients of powers of the various coupling strengths, then term by term  these coefficients are defined exactly, and they can be calculated exactly. But the summation of all these terms diverges, and only if one makes ``reasonable assumptions" about the final expressions, they can be compared with experiments. As long as the theoretical margins of error seem to be smaller than the experimental ones, theoreticians do not see the need for asking further questions. If the results agree, what more do you want? 
\itm{Two,} The Standard Model is a perfectly quantum mechanical one. There is, however, an on-going argument concerning the \emph{interpretation} of the quantum mechanical expressions. The theories we have, do not give infinitely precise predictions of the outcomes of experiments, but merely produce probabilistic distributions. Even if we had the \emph{exact} equations, without quantum mechanics at all, predictions for the experimental results would have had a stochastic nature anyway, since it is impossible to control exactly how two colliding particles will hit each other. \emph{The initial states are not exactly known}. Will the collision be a peripheral hit, or will it be head-on? The scattering parameters will always be unknown, so these always will have to be assumed to form flat distributions, which generates imperfections in the predictions that can be delivered. Why worry?
\ei
This author has been wondering how a perfect theory should be produced, or whether the basic shortcomings should be considered to be inevitable\,\cite{Jegerlehner.ref}\cite{GtHYang.ref}. Instead of making guesses as to the role of string-like elementary matter elements as opposed to point-like ones, or whether CFT/AdS techniques would cure the theories, we should perhaps try to construct models that explain how nature's degrees of freedom evolve, without any in-born imperfections. After all, we all agree that all particles generated in a collision experiment, are real, as opposed to probabilistic distributions. Can we do better?

\newsecl{Quantum Mechanics}{QM.sec}
In the 1920s, a group of physicists, in their discussions  at the Theoretical Physics 
Institute (later: ``Niels Bohr institute")  in Copenhagen, reached agreements as to what the theory of quantum mechanics says, and how to work with it. If you have the equations for the classical limit, 
you can almost \emph{derive} the quantum equations that would lead to that limit.

There was a more difficult issue: \emph{What is really going on, in a quantum system as we describe it?} 

Yet here also, general agreement was reached: It is amazing how well the theory predicts all probabilities without the need to answer this last question. The question cannot be answered by doing any experiment. Therefore, \emph{shut up and calculate.}\,\cite{Kampen.ref}

This came to be known as the Copenhagen Interpretation. It works perfectly, as if entirely correct.

But not all agree with the last dictum\,!

In our science, one is free to ask any question, so also this last one. It is legitimate to ask what it might be that is `truly happening'. Particles do not emerge in probabilistic distributions; there must be laws for their actual behavior as well.\cite{Bell.ref}\cite{Bell1982.ref}\cite{Bell1987.ref}.  By asking for these laws, we may learn more about our physical world. In particular, we wish to learn more about the why's of the Standard Model. Where do its fundamental constants come from?

The above question was asked, and discussed, for almost an entire century. \emph{God does not throw dice},  Perhaps He does throw dice, but He also calculates with infinite accuracy how the dice are going. Divergent perturbation expansions may be good enough to get an impression as to whether a theory may agree with experiment, but they should not be at the \emph{basis} of our theories. We should attempt to produce certain (``ontological") descriptions of what goes on. And we claim that such descriptions will be possible. Nothing besides our limited intelligence, will be standing in the way.

\subsection{Building models\labell{models.sub}}
	A good start could be that, where we used to focus on the `observer', we should rather begin with choosing a strategy for building \emph{models.}\,\cite{GtHFF.ref} We can demand for `ontological' models, models that really tell us what is going on. A typical example would be a \emph{cellular automaton}, a space-time consisting of cells.\cite{GtHCA.ref} In every cell we have discretised data (for instance data in the form of bounded integers, which would  allow us to sidestep the need for divergent perturbation expansions). The data evolve following exactly defined rules. These rules should not require any sort of superpositions.
	Suppose we first concentrate on small domains of a universe, temporarily bounding them by means of walls, so as to keep them totally finite. Being finite implies inevitably that the data reach values they possessed before, and hence such models are always periodic. It is not difficult to see that such models will always consist of periodic units. If we impose time reversibility (which would not be fundamentally necessary but simplifies things a lot), it is easy to see that the system consists entirely of periodic subunits. These subunits are all fundamentally of the same kind, so it is useful to consider these small, periodic systems as our primary ingredients for a useful theory. See Figure \ref{period.fig}.
	\begin{figure}
\qqquad\qqquad\widthfig{160pt}{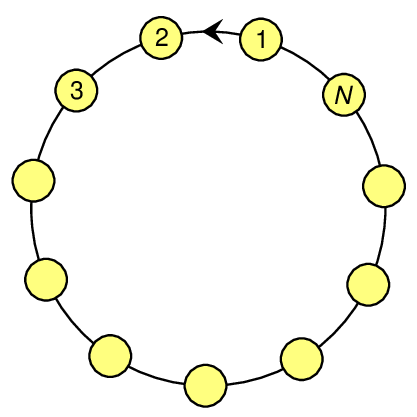}
\caption{The periodic chain. This example shows the case \(N=11\).} \labell{period.fig}
	\end{figure}
We can use the notation of quantum mechanics by representing the states this model can be in as Dirac kets. These kets can be regarded as an orthonormal set. The evolution law is presented as a chain, where we used our freedom to order the states: \(|0\ket\ra|1\ket\ra\cdots\ra|N-1\ket\ra|0\ket\)\,, see Fig.~\ref{period.fig}. In matrix notation:
\be |k\ket_{t+\del t}=U(\del t)|k\ket_t\ ,\qquad 
U(\del t)=\begin{pmatrix} 0&0&0&\cdots&1\\ 1&0&0&\cdots&0 \\ 0&1&0&\cdots\\[-3pt] \vdots && 1&\end{pmatrix}.\eel{classevolv}
We can find an operator \(H\) such that \(U(\del t)=\ex{-iH\,\del t}\) . The equation \(\frac{\dd|\psi\ket}{\dd t}=-iH|\psi\ket\), yields \(|\psi(t)\ket=|k\mod N\ket \) if \(t=k\,\del t\). and \(k\) is integer.
This definition of the Hamiltonian \(H\) is \emph{not unique}, but the following solution usually suffices.

The solution we shall use a lot, is obtained by the discrete Fourier transform:
\be |n^E\ket &\deff&\frac 1{\sqrt N}\sum_{k=0}^{N-1}\ex{2\pi i kn/N}|k\ket^\ontt\ , \qquad k=0,\cdots,\,N-1\ ; 
\labell{onttoE.eq}\\
|k^\ontt\ket      &=& \frac 1{\sqrt N}\sum_{n=0}^{N-1}\ex{-2\pi ikn/N}|n\ket^E\ , \qquad n=0,\cdots,\,N-1\ .\eel{Etoont.eq}
Here, the superscript `ont' refers to the ontological states \(|k\ket\) defined in Eq.\eqn{classevolv}, and the superscript \(E\) refers to the energy eigenstates of \(H\). Indeed, in the energy basis we find
\be H=\frac {2\pi}{N \,\del t}\,n=\w n\,,\qquad 0\le n<N\ .\eel{Heigenv.eq}  Clearly, these energy eigenstates have a regular sequence of eigenvalues listed by the integer \(n\).

In many applications, one can use the continuum limit of these equations. This is obtained by assuming \(\del t\) to be infinitesimal. In that case we write Eqs.\,\eqn{onttoE.eq}--\eqn{Etoont.eq} as
\be   |n\ket^E&\deff&\frac 1{\sqrt{2\pi}}\oint\dd\phi\,\ex{i\phi n/N}|\phi\ket^\ontt\ ,\qquad 0\le\phi<2\pi\,; 
\labell{contE.eq}\\
	|\phi\ket^\ontt&=&\frac 1{\sqrt{2\pi}}\sum_{n=0}^\infty \ex{-i\phi n/N}|n\ket^E\ . \eel{contont.eq}
We see that now the sequence of energy eigenvalues contains all integers \(n\) from 0 to \(\infty\).

Important theorem: at integer time steps, this Schr\"odinger equation sends collapsed wave functions (delta peaks) into collapsed wave functions. It does not generate superpositions (all states rotate along the circle with the same angular frequency \(\w\))

Important theorem: if system \(A\) has the same spectrum of energy eigenvalues as system \(B\), then a mapping \(A\leftrightarrow B\) exists; \,system \(A\) obeys the same evolution law as system \(B\), so that the two systems are physically identical. 

Note added: the converse is \emph{not} true; if two systems have different sets of energy eigenvalues, they may still describe the same ontological theory, as we can add integer multiples of 
\(2\pi/\del t\) to the energy levels of \(H\)  in Eq.~\eqn{Heigenv.eq}.

Note now that the energy spectrum of the ontological periodic system, \(E_n=\w n\,,\) is the same as the spectrum of a quantum harmonic oscillator.  Thus we have been describing quantum harmonic oscillators all the time. This implies that any system that consists of harmonic oscillators may be regarded as ontologically evolving classical systems.

\subsection{Bosonic field theories}\labell{freefields.sub}

An important example is the quantised field theory of free particles. These fields evolve as harmonic oscillators. They are the prototype of systems amenable to our procedure.
Consider a single, real particle (a particle is identical to its own antiparticle), as yet without interactions.  We consider the theory embedded in a box with size \(L\times L\times L\), obeying periodic boundary conditions at the walls. The Hamiltonian is 
		\be H=\int\dd^3\vec x\big(\half(\vec k^2+M^2)\Phi^2+\half\Pi^2\big)\ ,\ee
where the fields \(\Phi\) and \(\Pi\) obey local commutation rules, 
\([\Phi(\vec x,t),\,\Pi(\vec x\,'t']=i\del^3(\vec x-\vec x')\ ,\) and subsequently were Fourier transformed to \(\vec k\) space.

Because of the periodicity in \(\vec x\) space, \(\vec k\) space is discrete, quantised in units of size \(\del \vec k=2\pi/L\), while the values of the three \(k\) components range from \(-\infty\) to \(\infty\). 

We now apply the same transformations as those for the field \(\vv\) in Subsection~\ref{models.sub}, Eqs.~\eqn{contE.eq} and \eqn{contont.eq}, to arrive at an integer valued variable \(n(\vec k)\) at each value.\fn{There is a slight complication forcing us to first arrange the values of \(\vec k\) in pairs, \(\pm\vec k\), since \(\Phi\) and \(\Pi\) are \emph{real} fields in \(\vec x\) space; we skip such details for the time being.}  It is actually very important to postulate that values of \(\vec k\) in the far ultra-violet domain are kept out of the discussion, leaving more careful discussions for this feature for later (see Section~\ref{interact.sec}).

As is well-known in conventional quantum field theory, the integers \(n\) counts the particles  at each \(\vec k\) value , so that indeed the harmonic oscillator energy spectrum is arrived at, and we observe that the Fourier transformed variables \(\{\Phi(\vec k),\,\Pi(\vec k)\}\) represent motion in circles, with periodicities \(T(\vec k) =2\pi/\w(\vec k)\) where \(\w(\vec k)=\sqrt{M^2+\vec k^2}\).

\newsecl{Interactions}{interact.sec} In Ref\,\cite{GtHFF.ref} it is explained how any kind of `quantum interaction' can be approximated up to any desired accuracy, by modifying the evolution law in such a way that the interaction Hamiltonian \(H^\intt\) consists of pieces describing the various quantum transitions, multiplied by position operators that follow the motion \(t_i\) of the highest energy variables \(i\,\): 
	\be H^\intt=\sum_{\vec x,\,i}\half\pi\bpmatrix 0&-i\\ i&0\epmatrix\,\prod_i\del(t_i)\ . \eel{Hint.eq}
Here, for transparency, the integrals over \(\vec x\) were replaced by sums. The Pauli matrix \( \sig_2=\big(\scriptsize\bmatrix \, 0&-i\,\\ \, i&0\ematrix\big)\) stands short for all possible quantum flip operators for the slowest, explicitly observable particles. their explicit form depends on the basis chosen for the slowest variables\fn{There is some freedom in choosing this basis.} being the dynamical variables \(\Phi\) and \(\Pi\) at low values of \(|\vec k|\). Furthermore, \(t_i\) stand for the time dependence of the fastest degrees of freedom. Now, if we would ignore these latter factors, the coefficient \(\half\pi\) would guarantee an ontological evolution law, merely leading to flip operators for the interacting fields. In general, these flip-overs would happen too quickly, making the interaction Hamiltonian too strong. This is why we need the factors  \(\prod_i\del( t_i)\)\,. They add as limiting factor the probability that some fast variables all take a required prescribed value. Now, remember that we did not want to touch the highest energy components in momentum space. We now touch them as softly as possible, so that these high energies do come into play. The crucial step in the design of the quantum interaction effects in our models, is that the fastest variables are limited to the lowest energy states. All their excited states cost too much energy to be of relevance in accounting for the quantum states that will be considered. Limiting us to the ground states of the fastest variables also implies that all ontological data of the fast variables are distributed equally, so that they become basically invisible -- they are our invisible `hidden' variables.
  
  Notice, of course, that the introduction of new fast variables is not needed, they already exist in the form of the high-energy (virtual) particles. This forms the basic prescription for adding interactions among our physical degrees of freedom, which now cause explicit quantum effects: interactions can be computed by treating the interaction Hamiltonian Eq.~\eqn{Hint.eq} as a small perturbation:
  \be \bra\,\prod_i\del(t_i)\,\ket \equiv \eps\eel{intcoeff.eq}
must be small enough to compare the theory with existing quantum models, where the same expansion is used, apart from the fact that, in the usual quantised field theories no explanation is given as to where the value of \(\eps\) originated from. In our models we have discrete parameters \(N\) that may replace the continua in our description of bosons. This suggests the intriguing possibility that physical constants such as the finestructure constant \(\alf\) and the Kobayashi-Maskawa matrix, will eventually become computable in the distant future.

We have arrived at a theory where all actions and reactions are classical. There are no Hilbert space, no superpositions, no quantum entanglement (other than the entanglement of ordinary probability distributions). But the identification of the classical theories and the quantum ones, such as the Standard Model, only applies when we perform perturbation expansions, as we always do (indeed, we are forced to perform perturbation expansion to be able to do calculations at all). The question whether or to what extent these perturbation expansions converge or not, has become irrelevant, since the `real'  theory does not depend on perturbations at all, and only \emph{that} is our theory explaining the observed quantum effects.

\newsecl{A note on the Bell inequalities.}{Bell.sec}
In truly classical theories, such as ours, Bell's theorem does not hold. This is because the dynamical equations will never yield probability distributions unless distributions of the standard sort are \emph{already present} in the initial states. Indeed, beams of particles such as the ones used in particle colliders, already come as probability distributions. Even if in-going particles are as smooth and even as billiards, the results of multiple collisions would become basically unpredictable beyond some given probability distributions. 

It is the notion of \emph{statistical independence} that is violated in a maximal fashion: statistics only applies to theories that make approximations; without approximations, no statistical distributions. Not only would the outcomes of Alice and Bob's measurements be exactly determined by the equations, but also their choice of apparatus settings could not be controlled by free will\,\cite{CK.ref}, but by exactly valid equations.

We conclude that models can be constructed that explain quantum mechanical features in terms of exclusively classical equations of motion, without any need of non-locality or acausality.  In the case of repeated measurements, one must always reset the hidden variables before measuring again. 

`Counterfactual' observations are fundamentally impossible. In the author's opinion, the above observations are so evident that they should not need any further discussion. The weirdness we call quantum mechanics, and the fact the our world looks so different from a world with totally classical evolution laws, is  much more surprising. We have argued that this dissimilarity with the facts is delusive.

We do not claim that the exactly valid equations will be useful to mprove the outcome of Bell's thought experiment, since it will never be possible to register every single atom in the room, let alone in the minds of Alice and Bob. And there is no such thing as `free will'.

Investigations are continued, to obtain sharper terminology for explaining our findings.
\emph{Fermions} can be handled in a similar way, see Ref.\,\cite{Wetterich.ref}\cite{Wetterich2.ref}.

The author benefitted from many discussions, notably with T. Palmer, C. Wetterich, M. Welling and D. Dolce.

 \end{document}